\documentclass{article}

\usepackage{arxiv}

\usepackage[utf8]{inputenc} 
\usepackage[T1]{fontenc}    
\usepackage{hyperref}       
\usepackage{url}            
\usepackage{booktabs}       
\usepackage{amsfonts}       
\usepackage{nicefrac}       
\usepackage{microtype}      
\usepackage{lipsum}		
\usepackage{graphicx}
\usepackage{natbib}
\usepackage{doi}
\usepackage{amsmath}
\usepackage{amssymb}
\usepackage{xcolor}
\usepackage{subfigure}
\usepackage{subcaption}

\title{Data-Driven Discovery of Mobility Periodicity for Understanding Urban Systems}

\author{Xinyu Chen \\
	Massachusetts Institute of Technology \\
	\And
	Qi Wang \\
	Northeastern University \\
	\And
	Yunhan Zheng \\
	Massachusetts Institute of Technology \\
	\AND
	Nina Cao \\
	Massachusetts Institute of Technology \\
	\And
	HanQin Cai \\
	  University of Central Florida \\
	\And
	Jinhua Zhao\thanks{Corresponding author. Email: \texttt{jinhua@mit.edu}} \\
	Massachusetts Institute of Technology \\
}

\renewcommand{\shorttitle}{Data-Driven Discovery of Mobility Periodicity}

\hypersetup{
pdftitle={Data-Driven Discovery of Mobility Periodicity for Understanding Urban Systems},
}

\usepackage{lineno}

\begin{document}


\maketitle

\begin{abstract}

Human mobility regularity is crucial for understanding urban dynamics and informing decision-making processes. This study first quantifies the periodicity in complex human mobility data as a sparse identification of dominant positive auto-correlations in time series autoregression and then discovers periodic patterns. We apply the framework to large-scale metro passenger flow data in Hangzhou, China and multi-modal mobility data in New York City and Chicago, USA, revealing the interpretable weekly periodicity across different spatial locations over past several years. The analysis of ridesharing data from 2019 to 2024 demonstrates the disruptive impact of the pandemic on mobility regularity and the subsequent recovery trends. In 2024, the periodic mobility patterns of ridesharing, taxi, subway, and bikesharing in Manhattan uncover the regularity and variability of these travel modes. Our findings highlight the potential of interpretable machine learning to discover spatiotemporal mobility patterns and offer a valuable tool for understanding urban systems.

\end{abstract}

\keywords{Human mobility \and Transportation systems \and Regularity \and Periodicity \and Time series \and Machine learning \and Optimization}

\section{Introduction}

\begin{figure*}[bt]
\centering
\includegraphics[width=1\linewidth]{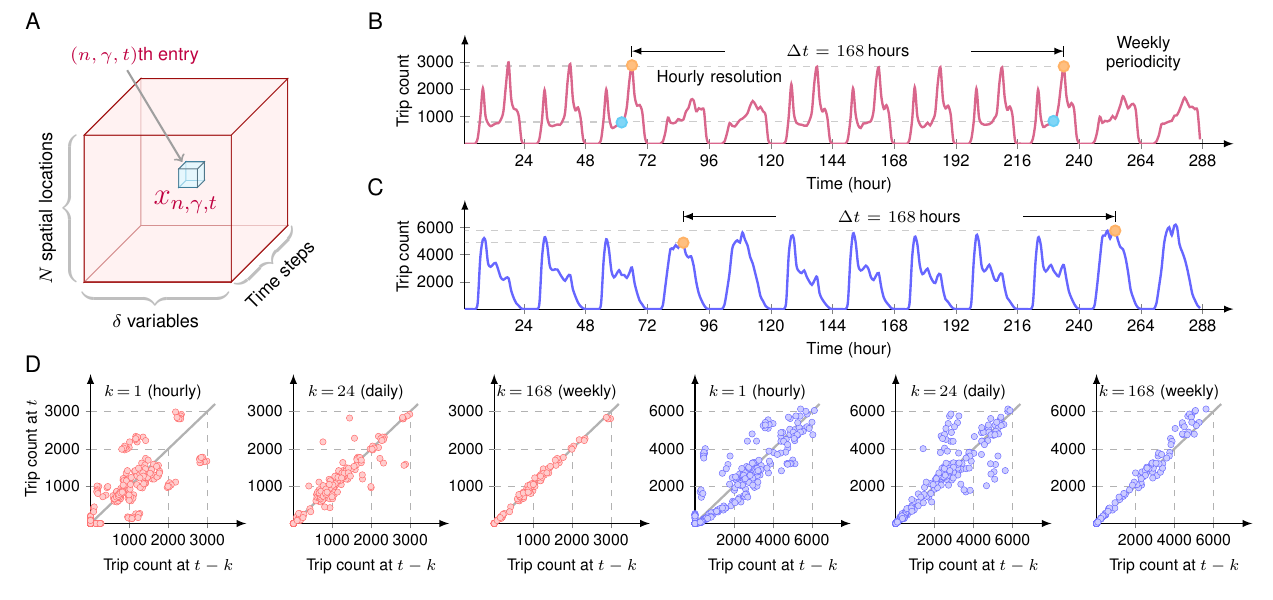}
\caption{Human mobility data and their time series periodicity. (A) Human mobility data can be structured as algebraic tensors where each entry corresponds to the value of a specific variable $\gamma$ at a particular spatial location $n$ and time step $t$. The variable dimension $\gamma$ is flexible and can encode different aspects of mobility, such as inflow and outflow at a particular spatial location. Moreover, the temporal lags such as data from different years can also be incorporated as distinct variables within this tensor framework. (B) Illustration of weekly periodicity in passenger flow time series. This panel displays an example of hourly passenger flow time series that exhibits a clear weekly periodicity. This periodicity is visually highlighted by annotating the data points according to weekly cycles (e.g., blue and orange circles). The time series reveals distinct mobility patterns between weekdays and weekends, with weekdays showing more pronounced morning and afternoon peak hours. (C) Example time series of passenger flow for comparison. (D) Auto-correlation analysis through scatter plots. These panels present scatter plots that visualize the relationship between trip counts at a given time step $t$ and trip counts at previous time steps $t-k$, in which $k$ represents different time lags (e.g., $k=1,24,168$ correspond to hourly, daily, and weekly cycles, respectively). Points that align closely along the anti-diagonal in these scatter plots implies positive auto-correlation and potential periodicity. Notably, the scatter plot for a weekly cycle demonstrates a more remarkable similarity compared to the daily and hourly cycles, visually confirming the strong weekly periodicity in the example time series.}
\label{metro_data}
\end{figure*}

Human mobility in cities often reveals recurring patterns---the predictable ebbs and flows of people moving through urban areas---that reflect the underlying structure of urban life \citep{gonzalez2008understanding, song2010limits, simini2012universal}. From daily commutes to weekend trips, these temporal regularities in human mobility underpin everything from urban transportation management \citep{sheffi1985urban} and transit scheduling \citep{guihaire2008transit} to predicting the spread of diseases \citep{balcan2009multiscale, belik2011natural, castillo2016perspectives, li2022spatiotemporal,du2018periodicity} and optimizing resource allocation \citep{katoh2024resource}. In essence, human mobility regularity is an important concept for identifying periodic trips in urban areas, which often follow daily and weekly cycles. A clear example is daily commuting, where passenger flows typically exhibit morning and afternoon peaks, forming a roughly ``M''-shaped pattern in trip time series data \citep{chen2022bayesian} (see e.g., Fig.~\ref{metro_data}B). As cities evolve, mobility patterns have become increasingly complex and less predictable, characterized by complicated regulatory policies, diverse transportation modes and behavioral shifts driven by social economic and environmental factors \citep{midgley2009role, machado2018overview, tirachini2020ride, nouvellet2021reduction, li2022spatiotemporal}. Thus, understanding the rhythm of human mobility in urban areas is essential for grasping how our societies function and urban systems operate \citep{gonzalez2008understanding, song2010limits, simini2012universal, batty2012smart, jiang2017activity, alessandretti2020scales}.

A paradigm for discovering long-term human mobility relies on collecting and constructing time series with a certain time resolution (e.g., hourly resolution) across different spatial locations over different years, which are often characterized as an algebraic tensor structure in Fig.~\ref{metro_data}A. In practice, tensors are multidimensional arrays which have been widely used to represent time series data of human mobility \citep{chen2022bayesian}. To explore human mobility regularity underlying these time series, a few critical but unsolved questions arise as follows. First, there is no standard method to quantify and compare the strength of mobility regularity and periodicity with different noise levels and temporal structures. The two example time series of passenger flow in Figs.~\ref{metro_data}B-C show different periodic patterns, while the strength of periodicity can be intuitively revealed by the auto-correlation analysis in Fig.~\ref{metro_data}D. Although it is obvious to observe from Fig.~\ref{metro_data}D that the weekly periodicity of Fig.~\ref{metro_data}B is stronger than Fig.~\ref{metro_data}C, the question that remains is how to compute the comparable metric for time series periodicity by referring to the auto-correlation plots in Fig.~\ref{metro_data}D. Second, there is no established approach to leverage periodicity strength as a metric for tracking the spatial and temporal evolution of mobility systems. Previous research has adopted methods such as Fourier Transform \citep{li2010mining,prabhala2014leveraging}, entropy-based measures \citep{goulet2017measuring,huang2019exploring,do2021estimating,teixeira2021impact}, reinforcement learning \citep{tao2021predicting}, and clustering of similar patterns \citep{manley2018spatiotemporal,zhong2015measuring, zhang2022periodicity} to identify periodicity and regularity in human mobility. In dynamical systems, dynamic mode decomposition identifies the cyclical patterns through the eigenvalue decomposition of vector autoregressive coefficients \citep{brunton2022data}.

However, these approaches typically do not provide an explicit and interpretable metric that directly quantifies periodicity, making them less suitable for downstream modeling or individualized interpretation. In addition, most studies that quantify periodicity or regularity in human mobility data focus on the analysis of individual trajectory data rather than aggregated demand or travel-mode usage data \citep{prabhala2014leveraging,goulet2017measuring,huang2019exploring,teixeira2021impact,do2021estimating,tao2021predicting,manley2018spatiotemporal,oliveira2016regularity}. While individual data can provide useful information on predicting mobility flows, interpretable information on aggregated demand data reveals longer term trends that can be more directly related to larger economic, social, and other changes that can help inform the development of policies in the future. In urban systems, human mobility patterns vary with different travel modes such as public transit, subway, ridesharing, and bikesharing. It is also meaningful to fairly quantify mobility regularity of different travel modes and utilize periodicity patterns to allocate resources and inform urban planning. Another fact is that mobility flows can be severely disrupted by extreme events and may evolve gradually over time due to changes in behavior, infrastructure, or policy. For example, the COVID-19 pandemic disrupted regular travel routines and led to a noticeable decline in mobility regularity \citep{li2022spatiotemporal, chen2025interpretable}. The challenge is how to systematically use periodicity strength to detect, quantify, and interpret such changes across both space and time.

Substantial progress has been made in understanding human mobility regularity and patterns from analytical perspectives such as entropy and predictability \citep{oliveira2016regularity, barbosa2018human, delussu2023limits}. However, the existing methods are not well-suited to quantifying periodicity across spatiotemporal domains and discovering dynamic patterns that related to human mobility regularity. Recently, interpretable machine learning methods \citep{murdoch2019definitions, rudin2022interpretable, brunton2022data} offer new opportunities to improve appearance, robustness, and interpretability of data-driven exploration in mobility data. In terms of model development, sparse linear regression \citep{tibshirani1996regression, jenatton2011structured} can be used to reformulate the problem of sparse identification of nonlinear dynamical systems \citep{brunton2016discovering}, which laid the foundations of several machine learning frameworks that designed for variable and feature selection in scientific discovery. In that venue, sparsity is one of the most important aspects to reinforce interpretability of machine learning models \citep{murdoch2019definitions, rudin2022interpretable}. However, existing sparse methods face critical limitations when applied to large-scale human mobility data. First, classical sparse regression methods are primarily designed for variable selection from univariate and multivariate data and struggle to scale to multidimensional data with spatial and temporal dependencies. Second, mobility data are inherently noisy, irregular, and heterogeneous across spatiotemporal domains, which challenges standard sparse models that lack built-in mechanisms for structured robustness. Third, even when sparsity is enforced, the resulting models do not yield interpretable or comparable metrics for periodicity strength across different spatial locations, variables, and time phases.

To bridge these gaps, we propose a multidimensional sparse autoregression framework that models time series of mobility flows across space and time, explicitly captures dominant auto-correlations, and provides a unified and interpretable measure of periodicity strength at daily and weekly cycles. This interpretable machine learning framework builds upon the tensor data structure in the field of multi-linear algebra. Formally, human mobility data are spatially- and temporally-varying, and can be represented as a third-order tensor of non-negative integers $x_{n,\gamma,t}\in\{0,1,2,\cdots\}$ as shown in Fig.~\ref{metro_data}A, where $n\in[N]$ indexes spatial locations, $\gamma \in [\delta]$ indexes different variables or time phases, and $t \in [T_\gamma]$ indexes time steps. This algebraic tensor structure captures the multidimensional nature of mobility systems and aligns with a rich body of work on tensorial machine learning for spatiotemporal pattern discovery \citep{kolda2009tensor, chen2024discovering}. Each individual sequence $x_{n,\gamma,t}$ is a time series, and thus can be seamlessly characterized by using time series models such as autoregression \citep{hamilton2020time}. As exemplified by Fig.~\ref{metro_data}B, real-world passenger flow time series demonstrate weekly periodicity by picking data points with a weekly cycle. When these data points are denoted by $x_t,t\in[T]$, the auto-correlations in Fig.~\ref{metro_data}D can be formulated as $x_t\approx w_kx_{t-k}$, measuring self-similarity in mobility over any time lag $k$. Observing Fig.~\ref{metro_data}D, data points of the example time series are very close to the anti-diagonal curve at a weekly cycle (i.e., $k=168$), implying that weekly periodicity is positively auto-correlated. Moreover, the ``closeness'' of data points to the anti-diagonal at index $k=168$ in Fig.~\ref{metro_data}D justifies the strength of weekly periodicity of time series, demonstrating that the time series in Fig.~\ref{metro_data}C is less periodic at a weekly cycle than Fig.~\ref{metro_data}B.

\section{Machine Learning Formula}

\begin{figure*}[bt]
\centering
\includegraphics[width=1\linewidth]{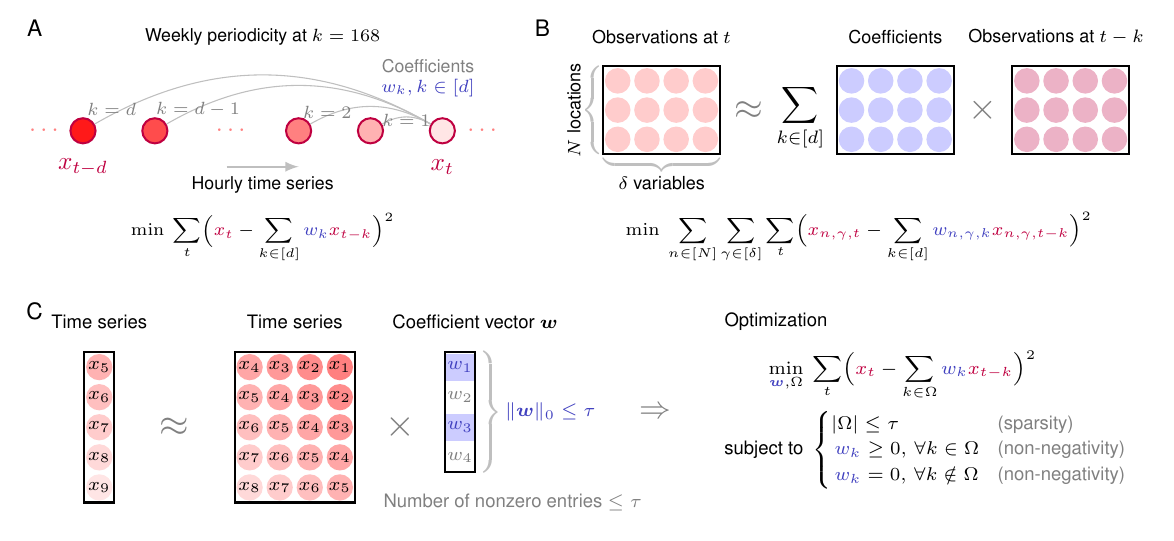}
\caption{Illustration of time series autoregression and its multidimensional setting on human mobility data. (A) The $d$th-order autoregression on the univariate time series can be formulated as an optimization problem for finding the optimal coefficients. Here, the order $d$ is an integer, determining the number of auto-correlations. (B) Multidimensional time series autoregression. Note that the symbol ``$\times$'' simply refers to the element-wise product between matrices. (C) The interpretability of time series autoregression can be reinforced by sparse and non-negative coefficients. The upper bound of the number of nonzero entries in the coefficient vector is set as an integer $\tau$, which is formally defined as the sparsity level. In mathematics, the cardinality of index set $\Omega$ (i.e., indices of nonzero coefficients) can be used to reformulate sparsity constraints. As a result, the decision variables in the constructed optimization problem include both coefficients and index set.}
\label{framework}
\end{figure*}

Interpretable machine learning provides a way to understand underlying patterns of complicated data and systems \citep{brunton2016discovering, murdoch2019definitions}. This study starts from the core idea of autoregression on time series, namely, the current value of the time series can be predicted as a linear combination of its past values \citep{hamilton2020time}. The objective is to find a set of coefficients that minimizes the sum of squared autoregressive errors (Fig.~\ref{framework}A) with predefined constraints such as sparsity and non-negativity on the coefficients as shown in Fig.~\ref{framework}C. Sparsity, enforced by the constraint $|\Omega|\leq\tau$ on the index set $\Omega$, limits the number of nonzero coefficients to a small integer $\tau$. This implies that only a few past time lags significantly contribute to predicting the current value. For instance, Fig.~\ref{metro_data}D shows that the trip counts at time $t$ are highly consistent with the trip counts at time $t-168$, revealing predictable and periodic patterns. By contrast, outliers on the scatter plots of Fig.~\ref{metro_data}D at the daily cycle, i.e., $k=24$, demonstrate less auto-correlated periodic patterns in the passenger flow time series as shown in Figs.~\ref{metro_data}B-C.

In the context of human mobility periodicity, we expect dominant nonzero coefficients that uncover periodic patterns such as a 168-hour cycle for weekly periodicity. Non-negativity of the coefficients ensures that the identified auto-correlations are positive, aligning with an intuitive understanding of periodicity as similar patterns repeating over time (see e.g., Fig.~\ref{metro_data}D). Multidimensional time series autoregression follows the specification of auto-correlations in which the coefficients over $N$ spatial locations and $\delta$ variables can be optimized (Fig.~\ref{framework}B). The optimization explicitly aims to find a small set of relevant past time lags in the index set $\Omega$ over $N\times \delta$ coefficient vectors. 
Formally, on the human mobility data $\{x_{n,\gamma,t}\}_{n\in[N],\gamma\in[\delta],t\in[T_{\gamma}]}$, we formulate the multidimensional sparse autoregression as follows,
\begin{equation}\label{sparse_ar_opt}
\begin{aligned}
\min_{\{{w}_{n,\gamma,k}\},\Omega}~&\sum_{n=1}^{N}\sum_{\gamma=1}^{\delta}\sum_{t=d+1}^{T_{\gamma}}\Bigl(x_{n,\gamma,t}-\sum_{k\in\Omega}w_{n,\gamma,k}x_{n,\gamma,t-k}\Bigr)^2 \\
\text{subject to}~&\begin{cases}
|\Omega|\leq\tau,\,\tau\in\mathbb{Z}^{+}, \\
w_{n,\gamma,k}\geq 0,\forall k\in\Omega, \\
w_{n,\gamma,k}= 0,\forall k\notin\Omega, \\
\sum_{k\in\Omega}w_{n,\gamma,k}=1, \\
\end{cases}
\end{aligned}
\end{equation}
where $w_{n,\gamma,k}, n\in[N],\gamma\in[\delta],k\in[d]$ are the non-negative coefficients of this $d$th-order autoregression formula, allowing one to quantify the positive auto-correlations such as daily and weekly periodicity as the dominant ones in time series. The last constraint is the normalization on the non-negative coefficients, letting the sum of the coefficients be one. Since the optimization problem is equivalent to the $\ell_0$-norm induced sparse autoregression, it can be addressed by mixed-integer optimization solvers with exact solutions \citep{bertsimas2016best, bertsimas2020sparse, bertsimas2024slowly, tillmann2024cardinality}. On such multidimensional tensor time series, we can discover the weekly periodicity of human mobility across different spatial locations and variables (or time phases).

On the hourly time series, if one intends to identify weekly periodicity, then the order of sparse autoregression should be set as $d\geq 168$ or greater. In what follows, we consider the order $d=168$ for the real-world mobility datasets. While the constraint $|\Omega|\leq\tau$ requires selecting a subset of $\tau$ coefficients out of $d$ possible ones, we extract the coefficients at index $k=168$, namely, auto-correlations $w_{n,\gamma,168},n\in[N],\gamma\in[\delta]$ to interpret the strengths of weekly periodicity across $N$ spatial locations over $\delta$ variables. The sparsity level $\tau$ is a critical hyperparameter for sparse autoregression. If this value is very small, then the index $k=168$ might not be in the support set $\Omega$, making it difficult to quantify weekly periodicity. In contrast, given a large sparsity level $\tau$, the values of auto-correlations within the support set $\Omega$ would have marginal differences.

\section{Metro Passenger Flow Periodicity}

To demonstrate the effectiveness of our framework in capturing periodicity from real-world mobility data, we begin with a case study of metro passenger flow data in Hangzhou, China. This dataset offers a clean and well-structured setting with strong known periodic patterns, making it an ideal testbed for validating the ability of the multidimensional sparse autoregression framework to recover meaningful temporal cycles---particularly weekly periodicity across different stations and flow directions. In addition, the optimized periodicity values of these passenger flow time series can be verified to be strictly consistent with auto-correlation analysis as shown in Fig.~\ref{metro_data}D.

\begin{figure*}[bt]
\centering
\includegraphics[width=1\linewidth]{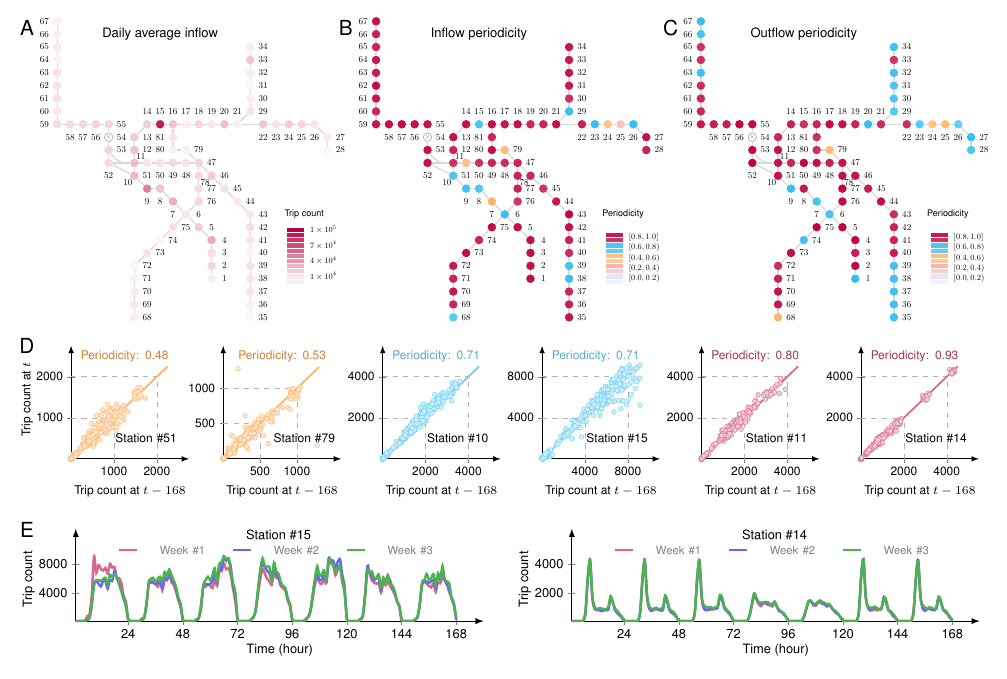}
\caption{Identification of weekly periodicity in inflow and outflow time series from the Hangzhou metro passenger flow dataset (from January 2 to 25, 2019) with the multidimensional sparse autoregression method (sparsity level $\tau=4$). (A) Daily average inflow trips of 81 metro stations. Note that the trip records of the station \#54 are missing. (B) Weekly periodicity of hourly inflow time series achieved by the sparse autoregression with sparsity level $\tau=4$. 64 stations show high periodicity ($\geq$ 0.8). The average periodicity value is 0.844. (C) Weekly periodicity of hourly outflow time series. 53 stations show high periodicity ($\geq$ 0.8). The average periodicity value is 0.829. Since inflow and outflow are two variables in the tensor data, their coefficients are optimized simultaneously with the same index set. (D) Scatter plots of trip counts at time $t$ versus time $t-168$ of 6 selected stations, illustrating varying strengths of weekly periodicity. The time series with a high weekly periodicity value also presents a strong similarity of data points between times $t$ and $t-168$. (E) Hourly inflow time series for stations \#14 and \#15, shown across three consecutive weeks: January 2–8, January 9–15, and January 16–22. Strong overlap among these weekly segments indicates high periodicity at a weekly cycle.}
\label{hangzhou_metro_periodicity}
\end{figure*}

The metro passenger flow dataset in Hangzhou\footnote{The anonymized trip records are available at \url{https://doi.org/10.5281/zenodo.3145404}.} was processed as a tensor structure with 81 stations, 2 variables (i.e., inflow and outflow), and 576 hours (i.e., over 24 days). Weekly periodicity is quantified by the non-negative coefficients at index $k=168$ in the multidimensional sparse autoregression with the prescribed order $d=168$. In the experiments, we set the sparsity level as $\tau=4$ depending on how many dominant auto-correlations we expect to identify. In the mixed-integer optimization solver, the support set $\Omega$ was optimized simultaneously with the coefficients. In this case, the support set is optimized as $\Omega=\{1,24,144,168\}$, corresponding to hourly, daily, 6-day, and weekly cycles, respectively. The coefficients at index $k=168$ refer to as the strength of weekly periodicity. The analysis of weekly periodicity of metro passenger flow reveals the mobility patterns of both inflow and outflow variables. Table~\ref{hangzhou_periodicity_tab} summarizes the detailed comparison of weekly periodicity between inflow and outflow variables. Observing Figs.~\ref{hangzhou_metro_periodicity}B-C and Table~\ref{hangzhou_periodicity_tab}, the inflow time series at the end stations of metro lines (e.g., $\{27,28\}$, $\{30,31,32,34\}$, $\{35,36,37\}$, and $\{63,66,67\}$) showed stronger periodicity than the outflow time series. Conversely, the outflow time series of stations in downtown areas (e.g., $\{7,8,10,15,51\}$) were found to be more periodic than the inflow time series. These findings are relevant to commuting trips in urban areas.

\begin{table}[ht!]
\caption{Strength of weekly periodicity of the Hangzhou metro passenger flow data at the selected end stations of metro lines. The weekly periodicity of inflow data in most selected stations is stronger than outflow data.}
\label{hangzhou_periodicity_tab}
\centering
\footnotesize
\begin{tabular}{c|cc||c|cc||c|cc}
\toprule
Station \# & Inflow & Outflow & Station \# & Inflow & Outflow & Station \# & Inflow & Outflow \\
\midrule
23 & \textbf{0.77} & 0.65 & 31 & \textbf{0.90} & 0.70 & 39 & 0.73 & \textbf{0.74} \\
24 & 0.46 & \textbf{0.48} & 32 & \textbf{0.91} & 0.74 & 40 & 0.83 & \textbf{0.88} \\
25 & 0.24 & \textbf{0.54} & 33 & \textbf{0.90} & 0.80 & 62 & \textbf{0.95} & 0.82 \\
26 & \textbf{0.78} & 0.62 & 34 & \textbf{0.90} & 0.76 & 63 & \textbf{0.89} & 0.77 \\
27 & \textbf{0.93} & 0.78 & 35 & \textbf{0.91} & 0.76 & 64 & \textbf{0.89} & 0.81 \\
28 & \textbf{0.87} & 0.69 & 36 & \textbf{0.88} & 0.74 & 65 & \textbf{0.86} & 0.80 \\
29 & \textbf{0.77} & 0.66 & 37 & \textbf{0.85} & 0.74 & 66 & \textbf{0.87} & 0.73 \\
30 & \textbf{0.81} & 0.71 & 38 & \textbf{0.76} & 0.75 & 67 & \textbf{0.93} & 0.79 \\
\bottomrule
\end{tabular}
\end{table}

Fig.~\ref{hangzhou_metro_periodicity}D shows the weekly periodicity values for six selected stations, measured as the strength of auto-correlations at a weekly lag (i.e., $k = 168$) in the inflow time series. The source emphasizes that time series with high weekly periodicity values also show a strong similarity between data points at times $t$ and $t-168$, indicating positive auto-correlations. The weekly periodicity is visually confirmed by scatter plots and time series overlaps. For instance, station \#14 shows a high weekly periodicity of 0.93, and its scatter plot shows a strong alignment along the anti-diagonal (Fig.~\ref{hangzhou_metro_periodicity}D). In particular, the comparison between stations \#14 and \#15 in Fig.~\ref{hangzhou_metro_periodicity}E answers why the strength of time series periodicity is a fair metric to reveal repeating patterns of human mobility. The periodicity value of station \#15 is 0.71, which is smaller than station \#14. As shown in Fig.~\ref{hangzhou_metro_periodicity}E, the trip count time series at station \#15 does not align  perfectly across weeks, indicating lower weekly periodicity compared to the more consistent pattern observed at station \#14. To summarize, on the passenger flow dataset, the comparable metric of weekly periodicity across different stations and inflow/outflow variables to discover periodic mobility patterns in a city-wide scale.



\section{Regularity of Ridesharing Trips}

\begin{figure*}[bt]
\centering
\includegraphics[width=1\linewidth]{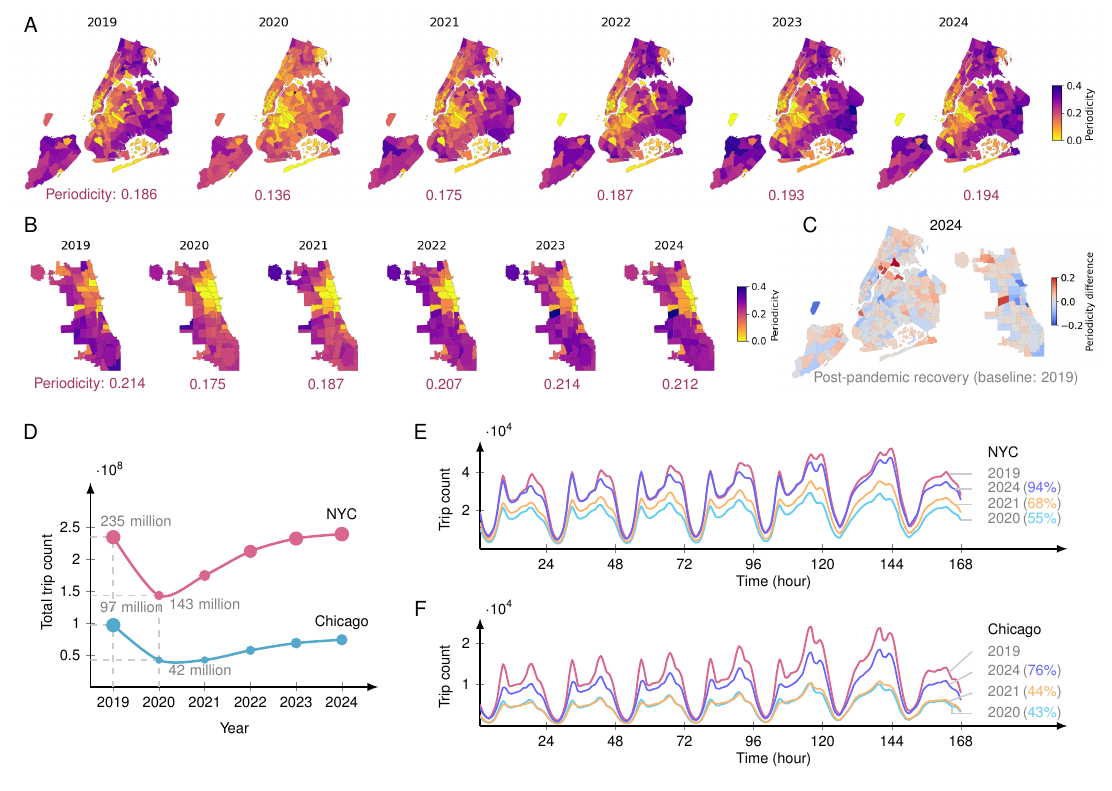}
\caption{Identification of weekly periodicity in ridesharing trip data across different urban areas with the multidimensional sparse autoregression framework at the sparsity level $\tau=6$. (A) Weekly periodicity of NYC ridesharing pickup trip data across 265 areas from 2019 to 2024. The average periodicity values are also marked. (B) Weekly periodicity of Chicago ridesharing pickup trip data across 77 areas from 2019 to 2024. (C) The coefficient difference between 2024 and 2019, referring to the coefficients of 2024 minus 2019. (D) Annual ridesharing trip counts of NYC and Chicago from 2019 to 2024. In contrast to NYC, the ridesharing trips in Chicago show a poor recovery in the post-pandemic years. Note that the ridesharing trip data of January 2019 is not available from the data portal. (E) Weekly average trip time series of NYC ridesharing, starting from Monday to Sunday. (F) Weekly average trip time series of Chicago ridesharing, starting from Monday to Sunday. While NYC shows the recovery percentage as 68\% in 2021, Chicago keeps 44\%, roughly as the same level of 2020.}
\label{rideshare_periodicity_pickup}
\end{figure*}

Multidimensional sparse autoregression quantifies human mobility periodicity through coefficients at regularity cycles. The resulting non-negative coefficients directly quantify the strength of the positive auto-correlation at certain time lags. A high coefficient at a time lag corresponding to a weekly cycle (e.g., 168 hours for hourly time series data) indicates a strong weekly periodicity. To analyze these periodicity values of different spatial locations over different years or different travel modes, the framework allows the extraction of spatial patterns that related to periodicity, the observations of long-term evolution of mobility regularity, and the discovery of changing behavior across different travel modes. As we know, the COVID-19 pandemic had a significant disruptive impact on urban mobility regularity and periodicity. To examine how such disruptions affected periodicity, we analyze large-scale ridesharing trip data from 2019 to 2024 in NYC\footnote{The NYC ridesharing trip record dataset is available at \url{https://www.nyc.gov/site/tlc/about/tlc-trip-record-data.page}.} and Chicago\footnote{The Chicago ridesharing trip record dataset is available at \url{https://data.cityofchicago.org/}.}. This analysis aims to demonstrate how the strength of weekly periodicity varied across different urban areas and years, reflecting the impact of the pandemic. In the experiments, we set the sparsity level and the order of sparse autoregression as $\tau=6$ and $d=168$, respectively. On both NYC and Chicago ridesharing datasets, the support set is optimized as $\Omega=\{1,23,24,143,167,168\}$, covering the indices of local auto-correlations, daily cycle, and weekly cycle.


In NYC, the weekly periodicity of pickup trips in 2019 was similar to the patterns observed in 2022--2024, indicating a return to pre-pandemic regularity after disruptions (Fig.~\ref{rideshare_periodicity_pickup}A). The year 2020 marked a significant shift from this regularity in NYC, with the pandemic severely disrupting trip patterns and reducing overall periodicity strength. By 2021, the ridesharing trips in NYC began recovering, transitioning from pandemic-induced irregularity towards the post-pandemic norm. During this recovery phase, suburban areas displayed increasingly periodic trip patterns in NYC. The periodicity patterns of ridesharing trips in 2022 became consistent with the patterns in 2019, while both periodicity values are almost same. Across these 6 years, ridesharing trips in downtown areas such as Manhattan were less periodic than suburban areas. In 2024, the overall weekly periodicity of NYC has been reinforced slightly. The ridesharing trips in suburban areas tend to be more periodic than 2019, while the downtown areas were less periodic (Fig.~\ref{rideshare_periodicity_pickup}C).

In Chicago, a similar pattern emerged where the overall periodicity of ridesharing trips in 2019 was stronger than in 2020, and downtown areas in 2020 were less periodic than in 2019 (Fig.~\ref{rideshare_periodicity_pickup}B). However, in contrast to NYC, the periodicity pattern transition in Chicago was slower, with no remarkable pattern change between 2020 and 2021. The periodicity patterns of ridesharing trips in Chicago during 2022--2024 also became consistent with patterns in 2019, suggesting a return to pre-pandemic regularity. In 2024, although the ridesharing trips demonstrated the same level of weekly periodicity as 2019, downtown areas became less periodic (Fig.~\ref{rideshare_periodicity_pickup}C). Comparing the aggregated trip counts in Fig.~\ref{rideshare_periodicity_pickup}D, NYC showed a quick return of ridesharing trip counts and reached the same level as 2019, while Chicago showed a poorer recovery in the post-pandemic years. The weekly average trip time series also indicated that while NYC showed a recovery percentage of 68\% in 2021 (Fig.~\ref{rideshare_periodicity_pickup}E), Chicago remained at 44\%, similar to its 2020 level (Fig.~\ref{rideshare_periodicity_pickup}F). Although Chicago showed a recovery percentage of 76\% ridesharing trips in 2024 (Fig.~\ref{rideshare_periodicity_pickup}F), the weekly periodicity already reached the same level as 2019.

Our analysis so far suggests that the COVID-19 pandemic led to a significant decrease in the periodicity of urban mobility in both NYC and Chicago in 2020. Both cities showed signs of recovery in subsequent years, with periodicity patterns returning to pre-pandemic levels by 2022--2024. However, the pace of recovery in terms of both periodicity and overall trip counts appeared to be faster in NYC compared to Chicago. Furthermore, our method enables policymakers to measure disruptions such as pandemics and design targeted recovery strategies.

\section{Mobility Periodicity of Multiple Travel Modes}

\begin{figure*}[bt]
\centering
\includegraphics[width=1\linewidth]{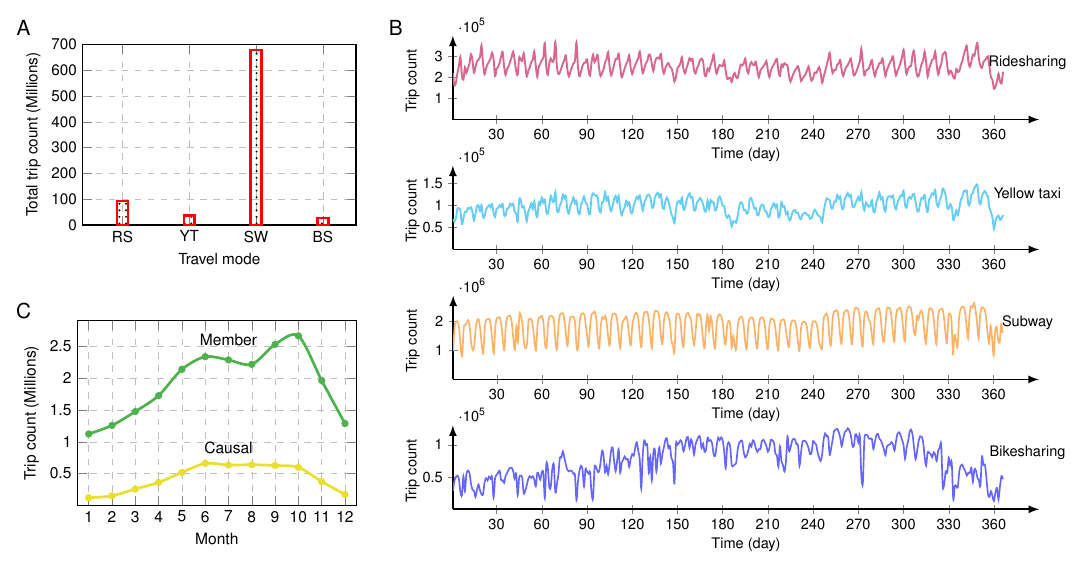}
\caption{Trip counts of ridesharing (RS), yellow taxi (YT), subway (SW), and bikesharing (BS) during the whole year of 2024 in Manhattan, one of the most densely populated boroughs of NYC. (A) The total trip counts of ridesharing, yellow taxi, subway, and bikesharing are 93.1 million, 36.4 million, 677.5 million, and 28.1 million, respectively. (B) Trip time series with a daily resolution in Manhattan. (C) Bikesharing trips of membership and causal users across 12 months of 2024. Most bikesharing trip records are related to the membership users.}
\label{nyc_time_series_manhattan}
\end{figure*}

In Manhattan, multi-modal trip records from 2024 reveal distinct weekly periodicity across four popular travel modes, including ridesharing, yellow taxi, subway, and bikesharing (Fig.~\ref{nyc_time_series_manhattan}). Through quantifying human mobility regularity with the multidimensional sparse autoregression framework, this analysis aims to demonstrate the differences of weekly periodicity across four travel modes and the underlying spatiotemporal patterns of multi-modal mobility data. As shown in Fig.~\ref{nyc_time_series_manhattan}A, total trip volumes decrease in the order of subway (677.5 million), ridesharing (93.1 million), yellow taxi (36.4 million), and bikesharing (28.1 million). The daily time series in Fig.~\ref{nyc_time_series_manhattan}B indicates that subway ridership exhibits the most remarkable periodicity, with highly regular weekly cycles sustained throughout the year. In contrast, ridesharing, yellow taxi, and bikesharing display lower periodic structures and greater short-term variability. Fig.~\ref{nyc_time_series_manhattan}C further disaggregates bikesharing trips into membership and casual users. Membership trips constitute 23.0 million rides, or approximately 82\% of the total, while casual trips account for the remaining 18\%. A strong seasonal pattern is evident, with trip counts peaking during the warmer months (May–October) and declining sharply in winter (January, February, December), reflecting the sensitivity of cycling activity to weather conditions.

\begin{figure*}[bt]
\centering
\includegraphics[width=1\linewidth]{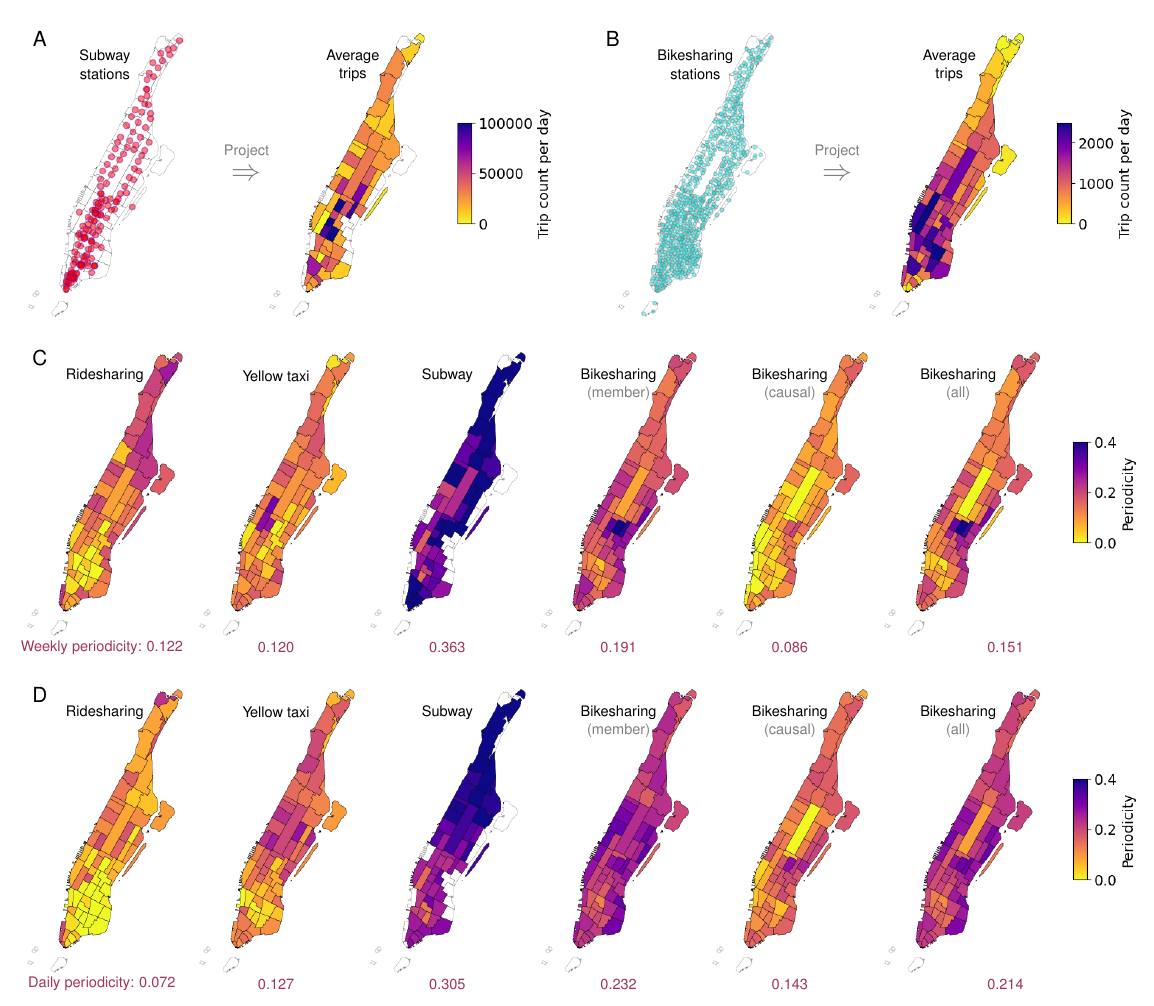}
\caption{Periodicity in multi-modal mobility trip data of 2024 in Manhattan with the multidimensional sparse autoregression method at sparsity level $\tau=6$. (A) Subway stations are projected onto 52 areas in Manhattan (i.e., 69 areas in total). (B) Bikesharing stations are projected onto 67 areas in Manhattan. (C) Weekly periodicity of four different travel modes, i.e., ridesharing, yellow taxi, subway, and bikesharing. According to the average periodicity, these travel modes are sorted as subway, bikesharing, ridesharing, and taxi in decreasing order. (D) Daily periodicity in weekday's multi-modal mobility trip data.}
\label{manhattan_result}
\end{figure*}

As shown in Fig.~\ref{manhattan_result}A–B, subway and bikesharing trip records are spatially projected onto spatial areas of Manhattan, yielding trip time series at the same spatial resolution as ridesharing and yellow taxi. Treating each travel mode as a variable, the hourly trip time series across spatial areas form a tensor (Fig.~\ref{metro_data}A). Given the distinct mobility patterns of bikesharing members and casual users, the variable dimension of mobility tensor comprises ridesharing, yellow taxi, subway, bikesharing (member), bikesharing (casual), and bikesharing (all). The resulting dataset is a tensor of size $69 \times 6 \times 8784$, with the temporal dimension spanning all 8,784 hours of 2024. By using the multidimensional sparse autoregression as described in Fig.~\ref{framework}, the support set is optimized as $\Omega=\{1,23,24,144,167,168\}$ where $k=168$ refers to as a weekly cycle. The auto-correlations at the index $k=168$ allow one to quantify the weekly periodicity of trip time series across different spatial areas and travel modes. Fig.~\ref{manhattan_result}C shows that subway trips are more periodic than other travel modes. Although the trips of ridesharing and yellow taxi have the same level of weekly periodicity, ridesharing trips are less periodic in the south areas and more periodic in the north areas of Manhattan than yellow taxi trips. In the bikesharing system, the comparison between membership and causal trips clearly demonstrates more periodic patterns of membership trips. As bikesharing trip volume is less than ridesharing and yellow taxi, the overall weekly periodicity of bikesharing is stronger than ridesharing and yellow taxi.

To examine time-varying mobility patterns, we analyze bimonthly trip datasets at an hourly resolution and quantify the weekly periodicity of different travel modes in Manhattan. As reported in Table~\ref{time_varying_periodicity}, weekly periodicity is generally lower in the early and late months of 2024 compared to the middle of the year. In particular, mobility regularity in November and December declines remarkably, possibly due to winter weather and holiday-related disruptions. While subway trips exhibit the strongest periodicity overall, they also experience a significant reduction during these months. By contrast, the trips of ridesharing and yellow taxi maintain relatively stable weekly periodicity throughout the year of 2024. Bikesharing usage demonstrates the greatest seasonal sensitivity, with notably weaker periodicity in colder months such as January–April and November–December.

\begin{table}[ht!]
\caption{Weekly periodicity of bimonthly trip data of four travel modes in Manhattan of 2024. The dataset was divided into 6 bimonthly time phases.}
\label{time_varying_periodicity}
\centering
\footnotesize
\begin{tabular}{c|cccc}
\toprule
& Ridesharing & Yellow taxi & Subway & Bikesharing (member) \\
\midrule
January \& February & 0.142 & 0.144 & 0.231 & 0.131 \\
March \& April & 0.191 & 0.198 & 0.449 & 0.134 \\
May \& June & 0.152 & 0.143 & 0.289 & 0.216 \\
July \& August & 0.183 & 0.133 & 0.408 & 0.219 \\
September \& October & 0.205 & 0.195 & 0.415 & 0.290 \\
November \& December & 0.046 & 0.055 & 0.136 & 0.120 \\
\bottomrule
\end{tabular}
\end{table}

In addition, we consider analyzing the daily periodicity of weekday's trips in Manhattan. The mobility tensor is of size $69\times 6\times 6288$ with the temporal dimension spanning 6,288 hours of 262 weekdays in 2024. The support set in the multidimensional sparse autoregression framework is optimized as $\Omega=\{1,23,24,95,96,120\}$ with sparsity level $\tau=6$ where the index $k=24$ refers to as a daily cycle. Observing the daily periodicity visualized in Fig.~\ref{manhattan_result}D, subway trips are more periodic than other travel modes, but the daily periodicity of the north areas is consistently stronger than the south areas of Manhattan. Compared to the consistent weekly periodicity of ridesharing and yellow taxi trips in Fig.~\ref{manhattan_result}C, the daily periodicity of yellow taxi trips is stronger than ridesharing as indicated by Fig.~\ref{manhattan_result}D. The daily periodicity of causal bikesharing trips is lower than membership trips, but the distinct regularity became a little marginal.


\section{Discussion}

Understanding the periodicity of human mobility is critical for revealing the underlying structure of individual and collective movement patterns in urban areas. Periodic patterns in human mobility—ranging from daily commutes to weekly rituals—are manifestations of deeply rooted behavioral rhythms. These patterns are essential components for making informed decisions, developing effective urban policies, and building more resilient and responsive urban systems. In literature, such rhythms have been extensively studied in chronobiology, behavioral sciences, and temporal network theory \citep{roenneberg2004marker,dijksterhuis2006making,holme2012temporal}, where periodicity reveals regularities in decision-making, attention cycles, and social synchronization. Drawing on these insights, we propose a novel method for quantifying and leveraging periodicity in large-scale mobility data. This enables a deeper understanding of urban rhythms and enhances predictive modeling for transportation planning, epidemic spread \citep{balcan2009multiscale,belik2011natural,castillo2016perspectives}, and urban resilience assessments \citep{xu2025using}.

Our results advance people's understanding of real-world urban systems and show the capability of multidimensional sparse autoregression for quantifying mobility regularity from time series data. Our findings concerning the weekly periodicity---a comparable metric across different time series, spatial areas, and time periods---in NYC and Chicago reveal the disruptive impact of the COVID-19 pandemic on mobility regularity and subsequent recovery trends. In Manhattan, the analysis of daily and weekly periodicity patterns of four different travel modes, including ridesharing, yellow taxi, subway, and bikesharing, demonstrates the rhythm and regularity strength across different spatial areas and bimonthly time phases. The interpretability offered by the multidimensional sparse autoregression framework is a key advantage, as sparse and non-negative auto-correlations provide actionable insights into the specific temporal lags driving the observed periodicity \citep{chen2025correlating}. The method we present allows substantial analysis of time series data that collected from real-world systems and demonstrated periodicity or seasonality. Despite identifying weekly periodicity, the revealed dominant auto-correlations in the support set enable us to introduce proper differencing operations---a simple yet effective solution to address the non-stationarity issue \citep{chen2025forecasting}.

As the field of human mobility research moves from analytical studies to data-driven experiments designed to directly learn underlying patterns, there is a growing need to formulate the periodicity quantification problem as a sparse identification of dominant positive coefficients in multidimensional time series autoregression. To this end, we develop an interpretable machine learning method for identifying spatial patterns and temporal dynamics that relate to weekly periodicity in urban areas. While this study focuses on metro passenger flow, ridesharing trips, and multi-modal travel behaviors in big cities, a potential limitation is how to extend these results to the cities of different scales and complicated transportation modes. Future work could integrate data from other transportation modes, such as public transit, to capture more comprehensive multi-modal behaviors. In terms of potential applications, this interpretable framework provides a robust foundation for discovering and quantifying temporal patterns related to periodicity and seasonality in complicated data and systems beyond human mobility, including climate systems, internet usage data, and web traffic time series.

\bibliographystyle{unsrtnat}
\bibliography{references}  






\clearpage

\appendix

\section{Data, Material, and Code Availability}

Data and code of this study will be available at \url{https://github.com/xinychen/integers}. For an intuitive understanding of time series periodicity, we also provide an interactive visualization tool on the ridesharing trip time series.

\section{Methodology}

While the optimization problem we have posed has the potential to learn a significant amount of information about periodicity in time series data, estimating the coefficient vectors $\{{w}_{n,\gamma,k}\}_{n\in[N],\gamma\in[\delta],k\in[d]}$ of $N$ spatial areas and $\delta$ phases when $N$ or $\delta$ become even moderately large becomes extremely computationally expensive. Thus, instead of estimating individual coefficient vectors $\{{w}_{n,\gamma,k}\}_{n\in[N],\gamma\in[\delta],k\in[d]}$, we first estimate global auto-correlations with sparsity constraints. The optimization problem for estimating a global coefficient vector $\boldsymbol{w}=\{w_k\}_{k\in[d]}\in\mathbb{R}^{d}$ from the human mobility data $\{x_{n,\gamma,t}\}_{n\in[N],\gamma\in[\delta],t\in[T_{\gamma}]}$ is given by
\begin{equation*}
\begin{aligned}
\min_{\{w_k\},\Omega}~&\sum_{n=1}^{N}\sum_{\gamma=1}^{\delta}\sum_{t=d+1}^{T_\gamma}\Bigl(x_{n,\gamma,t}-\sum_{k\in\Omega}w_{k}x_{n,\gamma,t-k}\Bigr)^2 \\
\text{subject to}&\begin{cases}
|\Omega|\leq\tau,\,\tau\in\mathbb{Z}^{+}, \\
w_{k}\geq 0,\,\forall k\in\Omega, \\
w_{k}=0,\,\forall k\notin\Omega, \\
\sum_{k\in\Omega}w_k=1,
\end{cases}
\end{aligned}
\end{equation*}
which can be solved by mixed-integer optimization. The support set is therefore optimized $\Omega$ simultaneously.

Given the global support set $\Omega$, learning the individual coefficients $\{w_{n,\gamma,k}\}_{n\in[N],\gamma\in[\delta],k\in[d]}$ can be simplified as follows,
\begin{equation*}
\begin{aligned}
\min_{\{w_{n,\gamma,k}\}}~&\sum_{t=d+1}^{T_{\gamma}}\Bigl(x_{n,\gamma,t}-\sum_{k\in\Omega}w_{n,\gamma,k}x_{n,\gamma,t-k}\Bigr)^2 \\
\text{subject to}~&\begin{cases}
w_k\geq 0,\,\forall k\in\Omega, \\
w_k=0,\,\forall k\notin\Omega, \\
\sum_{k\in\Omega}w_{k}=1,
\end{cases}
\end{aligned}
\end{equation*}
for any $n\in[N]$ and $\gamma\in[\delta]$. We solve this optimization problem by using quadratic optimization.

\section{Metro Passenger Flow Periodicity}

\subsection{Weekly periodicity on the 30-minute passenger flow time series data} 

\begin{figure}[ht!]
\centering
\subfigure[Inflow]{
    \centering
    \includegraphics[scale=0.6]{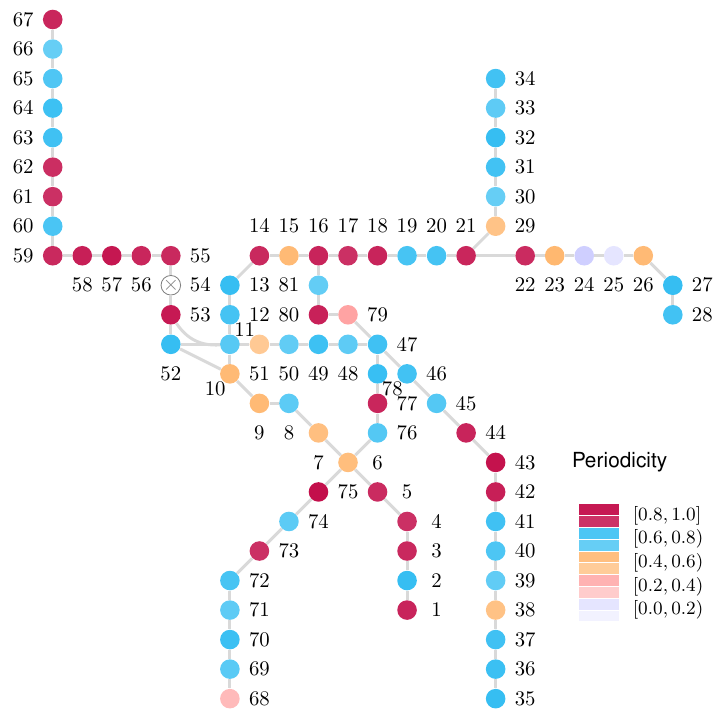}
}
\subfigure[Outflow]{
    \centering
    \includegraphics[scale=0.6]{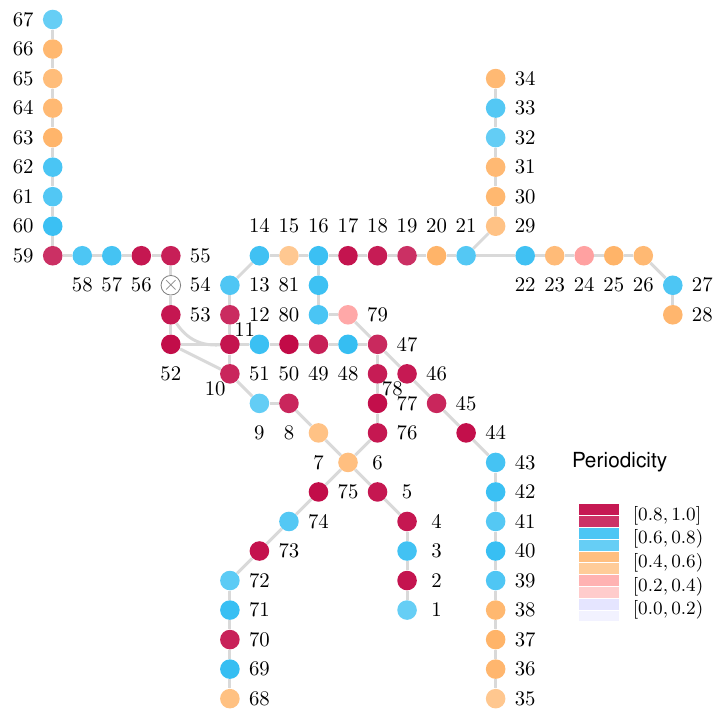}
}
\caption{Weekly periodicity of 30-minute inflow and outflow passenger flow in the Hangzhou metro system.}
\label{Hangzhou_k336}
\end{figure}

Fig.~\ref{Hangzhou_k336} shows the weekly periodicity across 80 stations by using the proposed model. While this study discusses passenger flow time series with both hourly and 30-minute time resolutions, it is possible to see the periodicity pattern shift of passenger flow from hourly data to 30-minute data. On these time series, the sparsity level of the proposed model is set as $\tau=4$, and the support set is optimized as $\Omega=\{1, 48, 335, 336\}$ (i.e., covering both daily cycle $k=48$ and weekly cycle $k=336$). Thus, the weekly periodicity is revealed by the auto-correlation at the index $k=336$. Comparing to the hourly time series, the weekly periodicity in Fig.~\ref{Hangzhou_k336} is relatively small because 30-minute time series demonstrate stronger local auto-correlations than hourly time series. Of the results, the inflow is more periodic than the outflow in the end stations, while the inflow is less periodic than the outflow in downtown areas. Thus, the time resolution impacts the time series periodicity quantification. Since the periodicity is a relative value, the comparison among time series with the same time resolution is still meaningful.

\section{Regularity of Ridesharing Trips}

\subsection{Weekly periodicity of ridesharing pickup trips with different sparsity levels}

\begin{figure}[ht!]
\centering
\includegraphics[width=0.95\textwidth]{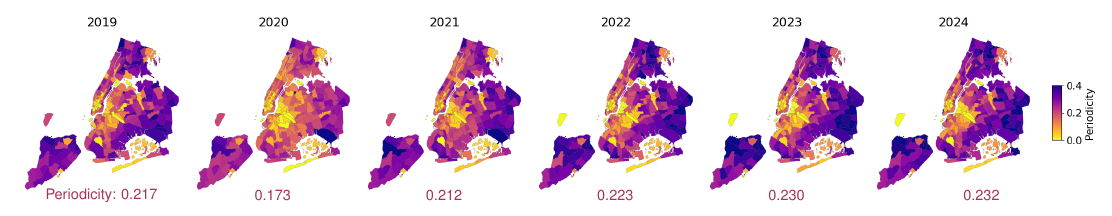}
\caption{Weekly periodicity of ridesharing pickup trips quantified by the proposed model with sparsity level $\tau=4$.}
\label{NYC_pickup_periodicity_tau}
\end{figure}

In this study, we consider the sparsity level on the NYC ridesharing data as $\tau=6$, covering weekly cycle in the support set. For comparisons, we set the sparsity level of the proposed model as $\tau=4$ in which the support set is optimized as $\Omega=\{1, 23, 167, 168\}$. As shown in Fig.~\ref{NYC_pickup_periodicity_tau}, although the weekly periodicity at the sparsity level $\tau=4$ is a little different from $\tau=6$, we can also observe the reduction of weekly periodicity in 2020 due to the COVID-19 pandemic, following by the recovery of human mobility regularity to the normal status. Thus, the interpretability of weekly periodicity is robust to the setting of different sparsity levels.

\subsection{Weekly periodicity of ridesharing dropoff trips with different sparsity levels}

In terms of dropoff trips, we set the sparsity levels of the proposed model as $\tau=4,6$ as follows.
\begin{itemize}
\item The support set with sparsity level $\tau=4$ is $\Omega=\{1,23,167,168\}$.
\item The support set with sparsity level $\tau=6$ is $\Omega=\{1,23,24,143,167,168\}$.
\end{itemize}

\begin{figure}[ht!]
\centering
\subfigure[Sparsity level $\tau=4$]{
    \centering
    \includegraphics[width=0.95\textwidth]{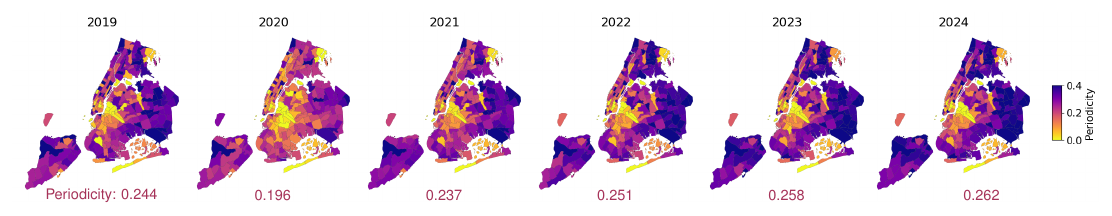}
}
\subfigure[Sparsity level $\tau=6$]{
    \centering
    \includegraphics[width=0.95\textwidth]{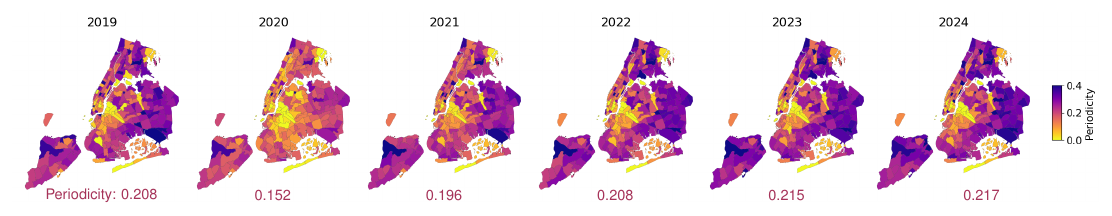}
}
\caption{Weekly periodicity of ridesharing dropoff trips quantified by the proposed model with sparsity levels $\tau=4,6$.}
\label{NYC_dropoff_periodicity_tau}
\end{figure}

As shown in Fig.~\ref{NYC_dropoff_periodicity_tau}, the dropoff trips of NYC ridesharing data demonstrated similar temporal patterns of weekly periodicity as the pickup trips. Trips with destination to downtown areas are less periodic than suburban areas. In 2020, a remarkable reduction of weekly periodicity can be clearly observed. In 2022, the weekly periodicity recovered to the same level as 2019. These findings intuitively verify the impact of the COVID-19 pandemic on mobility regularity in urban areas.

\end{document}